# A method to determine structural patterns of mechanical systems with impacts

*Barbara Blazejczyk-Okolewska*

*Division of Dynamics, Technical University of Lodz, Stefanowskiego 1/15, Lodz, Poland,*

**Abstract:**

A structural classification method of vibro-impact systems with an arbitrary finite number of degrees of freedom based on the principles given by Blazejczyk-Okolewska *et al*. [Blazejczyk-Okolewska B., Czolczynski K., Kapitaniak T., Classification principles of types of mechanical systems with impacts - fundamental assumptions and rules, European Journal of Mechanics A/Solids, 2004, 23, pp. 517-537] has been proposed. We provide a characterization of equivalent mechanical systems with impacts expressed in terms of a new matrix representation, introduced to formulate the notation of the relations occurring in the system. The developed identification and elimination procedures of equivalent systems and an identification procedure of connected systems enable determination of a set of all structural patterns of vibro-impact systems with an arbitrary finite number of degrees of freedom.

## 1. Introduction

In the theory of vibrations of mechanical systems, systems with impacts preoccupy an important place. They started to be investigated already in the mid 1950's and since then the interest in them has been growing and growing continuously (see, e.g., references [1-46]). Examples of their application include: physical models of buildings that are used to predict effects of earthquakes (e.g., Natsiavas [35], Nigm and Shabana [37]), pile-drivers for piles or pipes in oil mining, rammers for moulding mixes, crushers, riveting presses, hammer drills (e.g., Babitskii [2], Bajkowski [4], Kobrinskii and Kobrinskii [27]), vibration dampers (e.g., Bajkowski [4], Bapat [5], Mashri and Ibrahim [33], Peterka [39], Peterka and Blazejczyk-Okolewska [41]), low-loaded toothed and cam gears (e.g., Kaharaman and Singh [26], Lin and Bapat [29], Natsiavas [35], Nguyen *et al*. [36]), vibrating conveyors, bar screens, gun lock mechanisms, electric automatic cut-outs (e.g., Nguyen et al. [36]), printing heads in needle printers (e.g., Babitskii [2], Bapat [5], Kobrinskii and Kobrinskii [27], Tung and Shaw [45]), heat exchangers (e.g., Blazejczyk-Okolewska *et al*. [10], Lin and Bapat [29]). In numerous cases, e.g., in impact machines, vibration dampers or any other type shakers, this phenomenon (the phenomenon of impact) plays a very useful role. On the other hand, however, its occurrence is very undesirable, as it causes, e.g., additional dynamic loads, as well as faulty operation of machines and devices.

Intensive development of investigations on nonlinear phenomena comprises more and more complex vibro-impact systems. They differ as far as the design of their components is concerned, which results in various dynamical behaviour. While analyzing the studies devoted to mechanical systems with impacts, one can state that the researchers' attention has been drawn to systems that differ in (cf. [9]): i) a number $n$ of degrees of freedom (e.g., [13], [15], [29], [36], [38], [42]-[44], [46] for $n=1$, [2], [5]-[8], [13], [16], [28], [29] - $n=2$, [19], [29], [31] - $n=3$, [14], [31], [37] – arbitrary $n$), ii) a number $z$ of fenders (e.g., [16], [17], [24], [29], [43], [44] - $z=1$, [7], [8], [18], [29], [30], [34], [39], [41], [46] - $z=2$), iii) the way the limiting stops are arranged (e.g., [6], [7], [40], [41]), iv) designs of the supporting structure (e.g., [6], [7], [29], [46]), v) a number $e$ of excitations applied (e.g., [2], [5], [21], [23], [29], [33], [35]-[37], [41], [42], [45], [46] – $e=1$, [31], [40] - e>1 ).

During the investigations on dynamics of various mechanical systems, the author of the present study asked herself the following questions for many times: how a type of the system with impacts should be defined, how many such types can be differentiated and what their properties are. The comparative investigations of physical models of vibro-impact systems used in scientific studies have led to a presentation of assumptions and development of principles for the classification method of mechanical systems with one and two degrees of freedom (see Blazejczyk-Okolewska *et al.* [9]), which is shortly recalled in Section 2. Determination of all types of systems with impacts with one degree of freedom does not give rise to any difficulties. It is obvious that between the body and the frame there is only one supporting structure that can be described, for instance, with a certain function of displacement and velocity. However, in the case of systems with two degrees of freedom, a number of possible various combinations of connections and fenders (that describe the relations between two subsystems and the frame) grows significantly. It turns out that the determination of possible types of these systems in such a way as not to omit or repeat any of them is a much more difficult task to perform. Its solution for two, three and more degrees of freedom needs finding a proper method how to tackle this problem.

In the present study, a method for determination of all structural patterns of systems with impacts with an arbitrary finite number of degrees of freedom is discussed. The systems differ as far as the following issues are concerned: a number of degrees of freedom, a number and a configuration of fenders, and a number and a configuration of connections. To simplify the present considerations, it has been assumed that the possible connections are springs. Thus, to develop the proposed method, the following has been required: i) to use a matrix representation of the system with impacts, ii) to provide a characterization of equivalent systems according to the rules given in [9], iii) to develop procedures for generation of all possible combinations of these systems and to identify and eliminate unnecessary equivalent combinations, iv) to eliminate disconnected systems. The author would like to draw the readers' attention to the fact that the notion of equivalent mechanical systems is not identical with the notion of isomorphic systems (i.e., systems whose graphs are isomorphic), and, therefore, the standard methods for determining isomorphic graphs are not applicable here (see Subsection 4.1.2). The approach leads to an explicit division of all systems with impacts into disconnected subsets characterized by the fact that the behavior of systems of the same type (elements of one subset) can be described with equations of motion of the same structure.

The discussed classification of mechanical systems with impacts according to characteristic properties of their structure seems to be a natural classification. It reflects the relationships between the system structures, tells us about their way of evolution and presents their genesis. It allows us to rearrange the knowledge on systems with impacts and is the basis for understanding the sources of their diversity. Providing a full set of objects to be analyzed, it gives hints for new ideas and directions in designing technical devices.

## 2. Classification fundamental assumptions and principles

Let us recall the idea of the classification method proposed in [9]. Assume that the models of systems are rigid bodies with the masses $m_j$ ($j=1, 2,…, n$), connected by, e.g., springs, that can move along a straight line without a possibility to rotate. We say that a system has $n$ degrees of freedom if it is composed of $n$ bodies (referred to as subsystems further on) and it is not subdivided into independent systems. To simplify our considerations, the masses of elastic elements and the forces dissipating energy, except impact forces, will be neglected.

For a fixed number $n$ of degrees of freedom of the mechanical system, we build *the basic spring system* (with $s=[n(n+1)]/2$ springs), i.e., the system such that each subsystem (mass) is connected with another one and the frame by a spring, as well as *the basic impact system* (with $z=n(n+1)$ fenders), i.e., the system such that each subsystem impacts on any other subsystem and the frame at both possible senses of the relative velocity. If we remove even one spring from the basic spring system, we obtain another system (a system with another combination of spring arrangements). These systems will be referred to as *spring combinations*. The number of all possible spring combinations $i_s = 2^s$. Analogously, if we remove even one fender from the basic impact system, we obtain another system (a system with another combination of fender arrangements). These systems will be referred to as *impact combinations*. The number of all impact combinations $i_z = 2^z$. The basic spring systems and basic impact systems for *n=1,2,3* are given in Figs. 3-5 of [9].

Combining a basic spring system with a basic impact system, we get a *basic spring-impact system*, in which every subsystem is connected with any other subsystem and the frame and each subsystem impacts on any other subsystem and the frame at both possible senses of the relative velocity. Figure 6 of [9] shows basic spring-impact systems for one, two and three degrees of freedom. If even one spring or even one fender is removed from the basic spring-fender system, we obtain another system (a system with another combination of arrangements of springs or fenders). They will be referred to as *spring-impact combinations*.

All spring-impact systems are obtained as a result of the two-phase procedure (cf. [9]). *Phase I*, referred to as *a generation phase*, consists in matching each case of the spring combination with each case of the impact combination. It should be noticed here that the number of all possible spring-impact systems is considerably lower than the number $i_{sz} = 2^s \cdot 2^z = 2^{s+z}$ of all possible spring-impact combinations. Unfortunately, the applied method to differentiate spring-impact combinations has a certain fault. It turns out that combinations of models of spring-impact systems (with a different configuration of springs and fenders) looking apparently different can be assigned to the same physical model. Two examples of equivalent pairs of spring-impact systems are depicted in Figs. 7.b - 7.c of [9] and Figs. 7.d -7.e of [9]. It turns out that even for *n*=2 there are numerous spring-impact combinations which are equivalent to other combinations. An identification of the sets of equivalent combinations among spring-impact combinations (referred to as classes of relations) and a selection of their representatives will considered in Subsection 4.1.

All the systems in which a subdivision into two or more independent subsystems that are not connected either by a spring or by an impact occurs will be refereed to as *disconnected systems*. The systems in which a division into independent systems does not occur will be referred to as *connected systems*. A zone between subsystems which is defined by an action of at least one spring or fender connection will be called *the connectedness zone*. Let us notice that already for systems with *n*=3, matching a disconnected spring combination with a disconnected impact combination can lead to a connected spring-impact combination (see, e.g., the system analyzed by Dabrowski and Kapitaniak [19]). The identification of systems without zones of connectedness will be considered in Section 4.2.

The above-mentioned considerations lead to the second phase in determination of all spring-impact systems. *Phase II is an elimination phase* (cf. [9]) and it consists in elimination of redundant equivalent spring-impact combinations that correspond to one physical systems (subphase I) and to eliminate combinations that are faulty due to their disconnectedness (subphase II).

Here the author would like to point out that already for systems with *n*=2, identification of equivalent combinations is not a trivial task. If we consider a system with three, four or more degrees of freedom, we can state that it is difficult to control even the number of "subdivisions" into subsystems, not to mention the identification of equivalent combinations. An application of a matrix representation of the physical model of the mechanical system with impacts, proposed in Section 3, has contributed greatly to solving the above mentioned problems.

**3. Matrix representation of the physical model**

A natural way to describe numerous scientific and technical problems is modeling with graphs. As an example, one can mention here issues from the theory of switching and coding, analysis of electrical networks, operations research (including transport networks and game theory) and issues of program segmentation. It turns out that graphs obtained in practice for such problems are so large that their analysis without a computer is often impossible. Availability of fast computers has contributed undoubtedly to the current interest in graph theory.

Below a new way of representation of a physical model of the mechanical system with impacts is proposed. We will employ the graph terminology proposed by Deo [20].

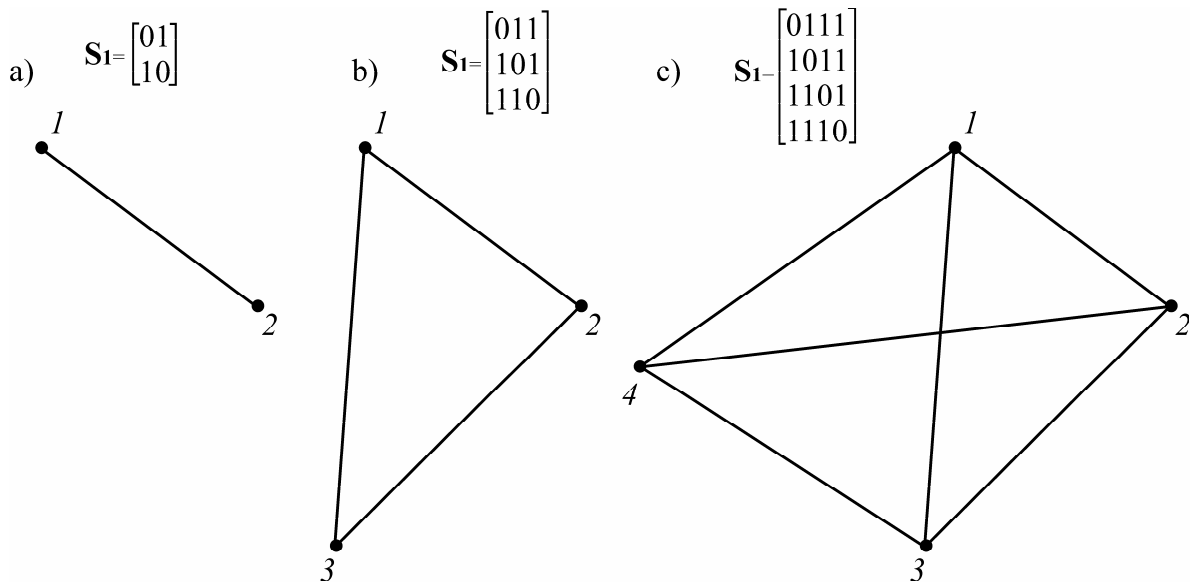

Fig. 1. Spring graphs for basic spring systems: a) n=1, b) n=2, c) n=3.

It has been assumed for the needs of the present study that the subsequent vertices will be the bodies of masses $m_j$, i.e., the subsequent subsystems *1, 2,…, n* up to the frame marked as *n+1*, and the edges – the segments that describe the connections between the subsystems, i.e., spring or impact connections. A graph which describes spring connections (spring relations) occurring in the system will be called *a spring graph*. The spring graph for a system with *n* degrees of freedom is always an undirected graph as each spring connects the subsystem with another subsystem or the subsystem with the frame (the frame with the subsystem) and the orientation of these connections is of no significance. Moreover, the spring graph does not have either parallel edges or self-loops (*a self-loop* is an edge whose ends are connected to one vertex). Spring graphs for basic spring systems of $n=1$, $n=2$ and $n=3$ have been shown in Fig. 1. They have the maximal number of edges (spring connections), in conformity with $i_s$ describing the number of springs. Figure 2.b illustrates a spring graph of the system shown in Fig. 2.a (hereafter, we follow the notation of springs and fenders used in [9]). One can read from it that there is a spring connection of subsystems *1* with *3* (subsystem *1* with the frame) and subsystem *2* with *3* (subsystem *2* with the frame), and that there is no spring connection between subsystems of the masses $m_1$ and $m_2$ (there is no edge connecting vertices *1* with *2*). This graph describes a disconnected spring combination.

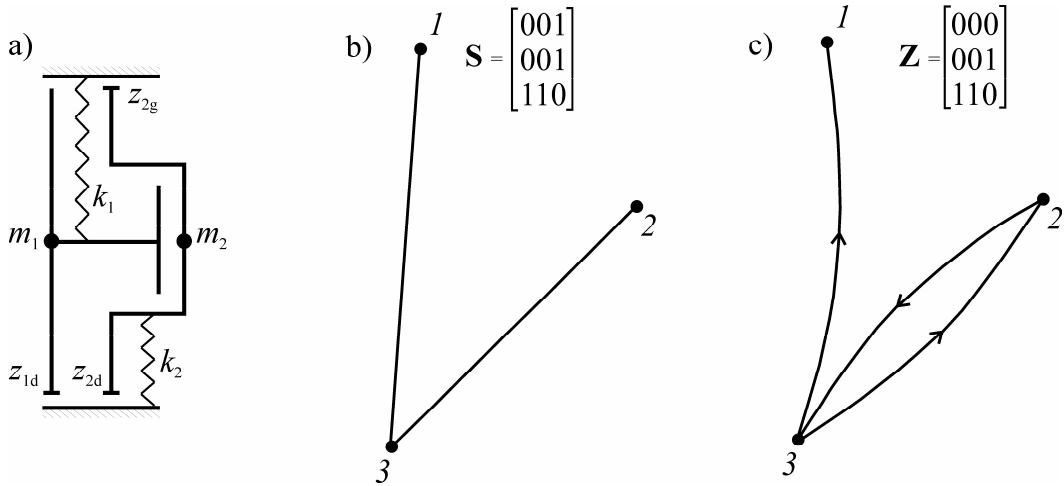

Fig. 2. Example of a disconnected spring-impact combination: a) scheme of the system, b) spring graph, c) impact graph.

A graph that describes impact connections (impact relations) occurring in the system will be referred to as *an impact graph*. A description of impact connections requires a sense of displacements of subsystems and the frame (although a displacement of the frame is not possible, we can imagine it for a while) to be accounted for. It leads to assigning proper directions (orientations) to the impact graph edges. While displacing each subsystem and the frame upwards (matter of convention), we encounter a fender of another subsystem or a fender of the frame on the way, then we can talk about an impact connection and we mark the edge orientation. Otherwise, there is no impact connection (there is no edge). While constructing a graph for a system with *n* degrees of freedom, we find that it will always be a directed graph with $n+1$ vertices that does not have any parallel edges or self-loops. As an example, let us analyze the impact graph from Fig. 2.c. There is an impact connection of frame *3* with subsystem *1* (the edge orientation informs us about it) in the graph, but not otherwise. There are also impact connections of *2* with *3* and of *3* with *2* (again, the respective orientation of the edge manifests it), and there are no impact connections of *1* with *2* and of *2* with *1*. This graph describes a disconnected impact combination.

Although a pictorial representation of the graph is very convenient and clear, a matrix representation is more suitable for computer processing. *An adjacency matrix of the undirected graph* (*spring graph*) with $n+1$ vertices and without parallel edges is called a symmetric binary matrix $\mathbf{S}=[s_{ij}]$ of the dimensions $n+1 \times n+1$ defined in such a way that $s_{ij}=1$, if there is an edge between the *i*th and the *j*th vertex, and $s_{ij}=0$, if there is no edge between them. The adjacency matrix $\mathbf{S}$ of the spring graph is called *a spring adjacency matrix*. Adjacency matrices of basic spring systems have been presented beside spring graphs in Fig. 1. At the spring graph in Fig. 2.b, its spring adjacency matrix $\mathbf{S}=[s_{ij}]$, *i, j*=1, 2, 3 has been written.

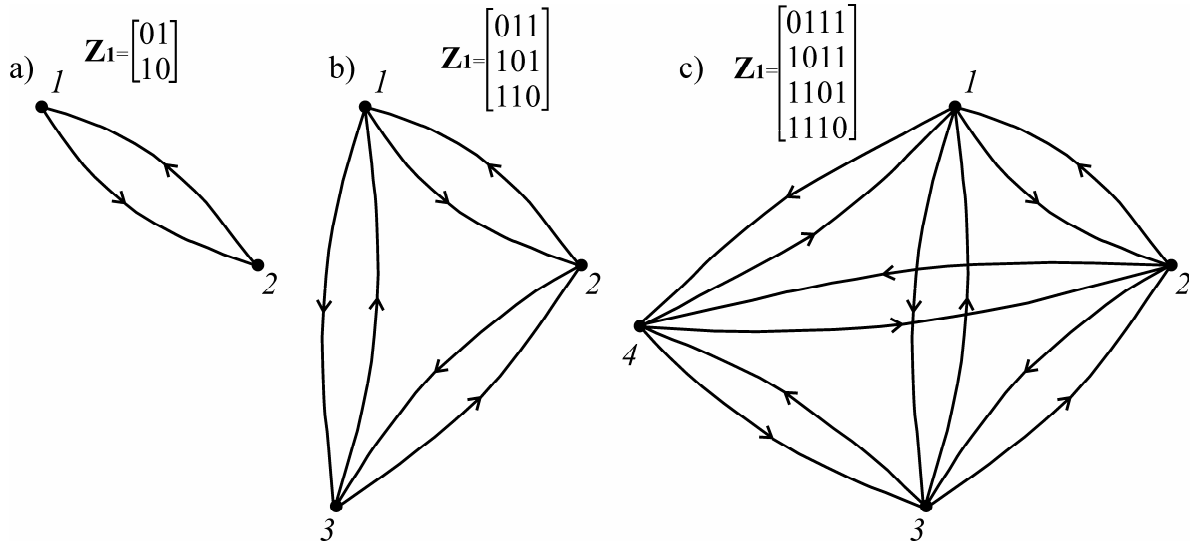

Fig. 3. Impact graphs for basic impact systems: a) n=1, b) n=2, c) n=3.

*An adjacency matrix of the directed graph* (*impact graph*) with $n+1$ vertices and without parallel edges is a 0-1 matrix $\mathbf{Z}=[z_{ij}]$ of the dimensions $n+1 \times n+1$ defined in such a way that $z_{ij}=1$, if there is an edge directed from the *i*th vertex to the *j*th vertex, and $z_{ij}=0$ otherwise. The adjacency matrix $\mathbf{Z}$ of the directed graph is called *an impact adjacency matrix*. At the impact graph in Fig. 2.c, there is an impact adjacency matrix $\mathbf{Z}$. Graphs of basic impact systems for n=1, 2, and 3 have been shown in Fig. 3a, Fig. 3b, and Fig. 3c. The impact adjacency matrix is symmetric if and only if an impact graph is *a symmetric directed graph* (it is such a graph in which for each edge from the vertex *a* to the vertex *b*, there is also an edge from the vertex *b* to the vertex *a*). A symmetric impact graph describes a symmetric physical system: if one subsystem impacts on another one via the upper fender, then it always has to be impacted on by the lower fender (see Fig. 3). An example of the unsymmetrical impact graph is a graph from Fig. 2.c, which describes impact connections of the system in Fig. 2.a.

By analogy, for the need of the present study, we can construct a spring-impact graph and introduce a notion of the adjacency matrix of the spring-impact system (a spring-impact adjacency matrix). A spring-impact system (e.g., this from Fig. 2.a) can be shown with two graphs, i.e., a spring graph (Fig. 2.b) and an impact graph (Fig. 2.c). The adjacency matrix of the spring-impact system (*a spring-impact adjacency matrix*) of the spring adjacency matrix **S** and the impact adjacency matrix **Z** is called a bloc matrix:

$$\mathbf{SZ} = [\mathbf{S}\ \mathbf{Z}]. \qquad (1)$$

The spring and impact graphs that describe spring-impact systems can be treated as one spring-impact graph and written on one drawing. Examples of spring-impact graphs for basic spring-impact systems of $n=1$, $n=2$ and $n=3$ have been shown in Fig. 4. They have the maximal number of edges (spring connections), described by the formula $i_s$, and the maximal number of directed edges (impact connections), given by the formula $i_z$.

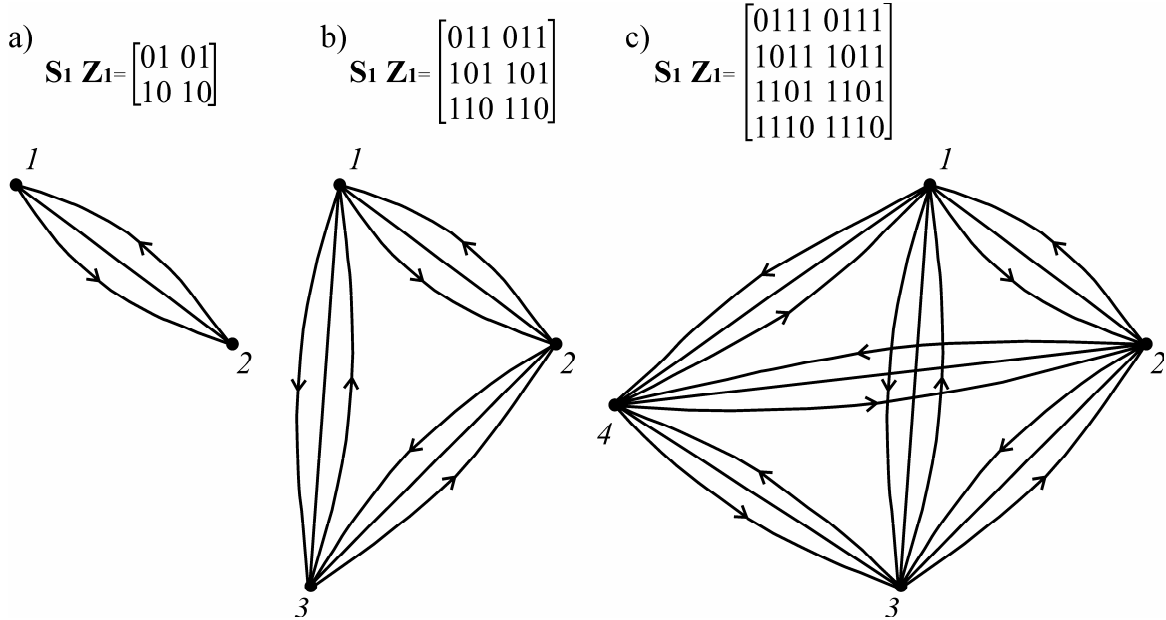

Fig. 4. Spring-impact graphs for basic systems and their adjacency matrices: a) n=1, b) n=2, c) n=3.

The above-described notions of the connectedness and the graph adjacency matrix can be a helpful tool to identify systems in which a subdivision into independent systems occurs – disconnected systems (subphase II of *Phase II*), and to identify the combinations equivalent to another combination (subphase I of *Phase II*).

**4. Classification method of spring-impact systems**

The majority of practical problems that can be modeled with graphs leads to so large graphs that they cannot be analyzed without a computer. This happens also in the case of issues considered in the present study. Of course, for spring-impact systems with a low number of degrees of freedom ($n=1$ or $n=2$), it is possible to draw a respective graph and state on this basis if it is connected or not. However, for higher $n$, investigations of properties (e.g., connectedness) of systems are not that easy. Firstly, a way in which subsequent spring-impact systems are generated should be determined in order not to omit or multiply any of them. The method that enables such generation of systems is presented in Subsection 4.1. In this Subsection, a characterization of equivalent systems is introduced and identification and elimination procedures of equivalent systems are given. In Subsection 4.2, a standard graph theory algorithm has been implemented for identification of connected systems. Subsection 4.3 includes the final procedure for classification of systems.

## 4. 1. Generation of spring-impact systems

### 4. 1. 1. Generation of all adjacency matrices

Spring and impact adjacency matrices are binary matrices with elements equal to "0" or "1". To determine a set of all such matrices, we will use a representation of natural numbers in the binary system. Let us remind here that in this system the expression $c_n c_{n-1} \ldots c_0$, where $c_n, c_{n-1}, \ldots, c_0$ are 0's or 1's, denotes the number $c_n 2^n + c_{n-1} 2^{n-1} + \ldots + c_0 2^0$.

*Description of the procedure of constructing spring and impact combinations with n degrees of freedom:* Constructing an adjacency matrix of spring graphs and an adjacency matrix of impact graphs consists in a generation of respective binary series and a proper arrangement of their terms in the matrix tables. A simple way to find successive digits of the binary notation of the number presented in the decimal notation is finding remainders of subsequent divisions by two for a series of numbers where the first term is the number whose binary expression we seek for and the next terms are integral parts from the previous divisions. Reversing the sequence of the terms in the obtained series of remainders, we obtain the binary expansion sought. A number of series corresponding to all spring adjacency matrices for $n$ degrees of freedom is equal to $2^s$. As each adjacency matrix of the spring graph is symmetric, it is enough to generate the respective triangular matrix. A number of series corresponding to all impact adjacency matrices for $n$ degrees of freedom is equal to $2^z$. The adjacency matrix of the impact graph does not have to be symmetric, thus we have to have $2^z$ of binary series.

The matrix representation proposed in this study allows us to investigate connectedness. Its disadvantage lies in the fact that various adjacency matrices can correspond to one physical system. This fault can be overcome via identification and elimination of unnecessary matrices.

Having in mind convenience of this presentation, we will refer to spring, impact and spring-impact combinations as spring, impact and spring-impact systems to the end of Subsection 4.1.

## 4. 1. 2. Characterization and identification of equivalent systems

Subphase I of *Phase II* (Section 2) comprises elimination of spring-impact systems equivalent to another system.

To characterize and identify equivalent systems, let us introduce the following notions: a transposed adjacency matrix, an inversed adjacency matrix and a translocated adjacency matrix.

A transposed matrix to the adjacency matrix **A** is called *a transposed adjacency matrix* and is denoted with the symbol $\mathbf{A}^T$. The system described by $\mathbf{A}^T$ will be referred to as the transposed one and the operation due to which we obtain this system – a transposition of the system. The transposition of the system (the adjacency matrix of the system) can be treated as a change in the orientation of the frame of reference introduced during the investigations of the system dynamics.

Let **A** be an adjacency matrix of the type $n+1 \times n+1$. Inversing rows $w_i$ with rows $w_{n-i+1}$ for $i=1, 2,…, [(n+1)/2]$, and then inversing columns $k_j$ with columns $k_{n-j+1}$ for $j=1, 2,…, [(n+1)/2]$, we obtain a new adjacency matrix $\mathbf{A}^P$, which will be called *an inversed adjacency matrix*. The symbol $[x]$ denotes the integral part of the number $x$. The system described by $\mathbf{A}^P$ will be called an inversed system and the operation due to which we obtain this system – an inversion of the system. The inversion of the system causes a change in the arrangement of vertices (as a matter of fact, a change in the arrangement of subsystems).

The matrix transposed to the inversed adjacency matrix $\mathbf{A}^P$ is called *a translocated adjacency matrix* $\mathbf{A}^{PT}$. The system described by $\mathbf{A}^{PT}$ will be called a translocated system, and the operation due to which we obtain this system – a translocation of the system. The translocation of the system causes a change in the arrangement of vertices (as a matter of fact, in the arrangement of subsystems), and then a change in the orientation of edges (change in the orientation of the frame of reference).

The above-mentioned definitions concern the spring systems ($\mathbf{S}^T$, $\mathbf{S}^P$, $\mathbf{S}^{PT}$), the impact systems ($\mathbf{Z}^T$, $\mathbf{Z}^P$, $\mathbf{Z}^{PT}$), and the spring-impact systems ($\mathbf{SZ}^T=[\mathbf{S}^T\mathbf{Z}^T]$, $\mathbf{SZ}^P=[\mathbf{S}^P\mathbf{Z}^P]$, $\mathbf{SZ}^{PT}=[\mathbf{S}^{PT}\mathbf{Z}^{PT}]$).

Let **A** and **A**′ be spring or impact adjacency matrices, respectively. We say that the system **A** is *equivalent via transposition, inversion, or translocation* with the system **A**′, when $\mathbf{A}^T=\mathbf{A}'$, $\mathbf{A}^P=\mathbf{A}'$, $\mathbf{A}^{PT}=\mathbf{A}'$, correspondingly. If at least one of these equivalencies holds, we can say that the systems **A** is *equivalent* to **A**′.

The identification of spring-impact systems equivalent to other spring-impact systems is conducted in three different ways.

We say that *a transposed equivalence* (way I) *of the spring-impact system* [**SZ**] to the spring-impact [**S**′**Z**′] occurs when:

$$\mathbf{S} = \mathbf{S}' \wedge \mathbf{Z}^{\mathrm{T}} = \mathbf{Z}'. \tag{2}$$

The symbol $\wedge$ denotes a conjunction. The spring-impact system [**SZ**] is *equivalent to itself via transposition* if $\mathbf{SZ}^{\mathrm{T}} = \mathbf{SZ}$.

We say that *an inversed equivalence* (way II) *of the spring-impact system* [**SZ**] to the spring-impact system [**S′Z′**] occurs when:

$$\mathbf{S}^{\mathrm{P}} = \mathbf{S} = \mathbf{S}' \wedge \mathbf{Z}^{\mathrm{P}} = \mathbf{Z}' \tag{3a}$$

or

$$\mathbf{S}^{\mathrm{P}} = \mathbf{S}' \wedge \mathbf{Z}^{\mathrm{P}} = \mathbf{Z}' \tag{3b}$$

or

$$\mathbf{S}^{\mathrm{P}} = \mathbf{S}' \wedge \mathbf{Z}^{\mathrm{P}} = \mathbf{Z} = \mathbf{Z}'. \tag{3c}$$

The spring-impact system [**SZ**] is *equivalent to itself via inversion* if $\mathbf{SZ}^{\mathrm{P}} = \mathbf{SZ}$.

We say that *a translocated equivalence* (way III) *of the spring-impact system* [**SZ**] to the spring impact system [**S′Z′**] occurs when:

$$\mathbf{S}^{\mathrm{P}} = \mathbf{S} = \mathbf{S}' \wedge \mathbf{Z}^{\mathrm{PT}} = \mathbf{Z}' \tag{4a}$$

or

$$\mathbf{S}^{\mathrm{P}} = \mathbf{S}' \wedge \mathbf{Z}^{\mathrm{PT}} = \mathbf{Z}' \tag{4b}$$

or

$$\mathbf{S}^{\mathrm{P}} = \mathbf{S}' \wedge \mathbf{Z}^{\mathrm{PT}} = \mathbf{Z} = \mathbf{Z}'. \tag{4c}$$

The spring-impact system [**SZ**] is *equivalent to itself via translocation* if $\mathbf{SZ}^{\mathrm{PT}} = \mathbf{SZ}$.

We say that the spring-impact system [**SZ**] *is equivalent* to the spring-impact system, [**S′Z′**] if a transposed equivalence or an inversed equivalence takes place or a translocated equivalency occurs. Then, we write:

$$[\mathbf{SZ}] \sim [\mathbf{S}' \, \mathbf{Z}']. \tag{5}$$

The following conclusions result from the above-mentioned definitions, namely:

1) We trend to identify spring-impact systems that are assigned to one model but while generating all possible combinations (*Phase I* – generation phase), they were treated as different models. It has been observed that the system in which the following was altered: a) the orientation of the frame of reference (2), b) the sequence of subsystem numeration (3), c) both the sequence of subsystem numeration as well as the orientation of the frame of reference (4) became a new model in an artificial way.

2) Transposition of the system (the adjacency matrix of the system) can be treated as a change in the orientation of the frame of reference introduced during the investigations of the system dynamics. As the spring adjacency matrix is symmetric, thus it is equivalent (via transposition) to itself only. Transposing the impact adjacency matrix, we obtain either the original matrix (the matrix is equivalent to itself via transposition) or the equivalent matrix via transposition with the original matrix. The first case occurs when the impact adjacency matrix is symmetric (for each edge from the vertex *a* to the vertex *b*, there is an edge from the vertex *b* to the vertex *a*). The second case takes place when both impact graphs have edges oriented in opposite directions.

3) Inversion of the system, i.e., inversion of rows and the corresponding columns of spring adjacency matrices or impact adjacency matrices (see the above definition of the inversed adjacency matrix) causes a change in the arrangement of vertices (as a matter of fact, a change in the arrangement of subsystems). However, it should be noticed that rows and columns are to be arranged in the same order. Thus, if two matrix rows are interchanged (e.g., S or Z), then the columns corresponding to them should be interchanged as well. The inversion of spring adjacency matrices or impact adjacency matrices leads to obtaining: either the original matrix (the matrix equivalent to itself via inversion) or the matrix equivalent to the original matrix (via inversion).

4) Translocation of the system, that is to say, an inversion of rows and corresponding columns in the spring adjacency matrix or the impact adjacency matrix, and then their transposition causes a change in the arrangement of vertices (as a matter of fact, in the arrangement of subsystems), and then a change in the orientation of edges (change in the orientation of the frame of reference). Hence, the properties the translocated matrices will be characterized by comply with the description of points 2 and 3. The translocation of spring or impact adjacency matrices does not change either the original matrix (the matrix is equivalent to itself via translocation) or leads to obtaining a matrix equivalent to the original one via translocation. The procedure of translocation of the given spring system is identical to the procedure of inversion of this system. This is not always the case for impact systems, however.

The spring-impact systems from Fig. 7b and Fig. 7c of [9], Fig. 7d and Fig. 7e of [9], and $[S_2Z_5]$ and $[S_5Z_2]$ with such $S_2$, $S_5$, $Z_2$, $Z_5$ as in Fig. 5a, Fig. 5b, Fig. 6a, Fig. 6b, are examples of systems in which equivalency, respectively via transposition (way I), via translocation (way III), and via inversion (way II) take place.

If two spring-impact systems are equivalent (via transposition, inversion or translocation), then we say that they belong to the same class of relations. Elements of the class of relations have to be identified with an arbitrary chosen representative of this class.

It should be pointed out that there exist nonequivalent vibro-impact systems whose graphs are isomorphic and therefore the standard methods for determining isomorphic graphs are not applicable here. For example, the impact systems with adjacency matrices $\mathbf{Z}=[z_{ij}]$ and $\mathbf{Z'}=[z'_{ij}]$, *i*, $j$=1, 2, 3, where $z_{ij}$=0 if ($i=j$) or ($i$=1 and $j$=2) and $z_{ij}$=1 otherwise, and $z'_{ij}$=0 if ($i=j$) or ($i$=3 and $j$=1) and $z'_{ij}$=1 otherwise, have isomorphic graphs, but they are not equivalent.

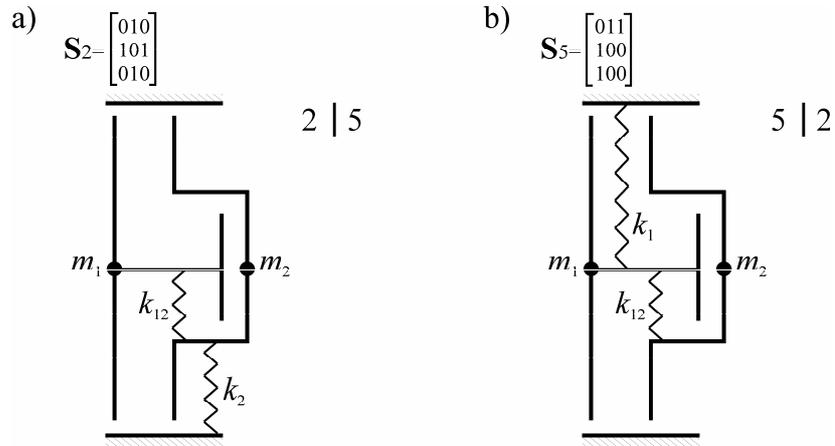

Fig. 5. Example of equivalent spring systems.

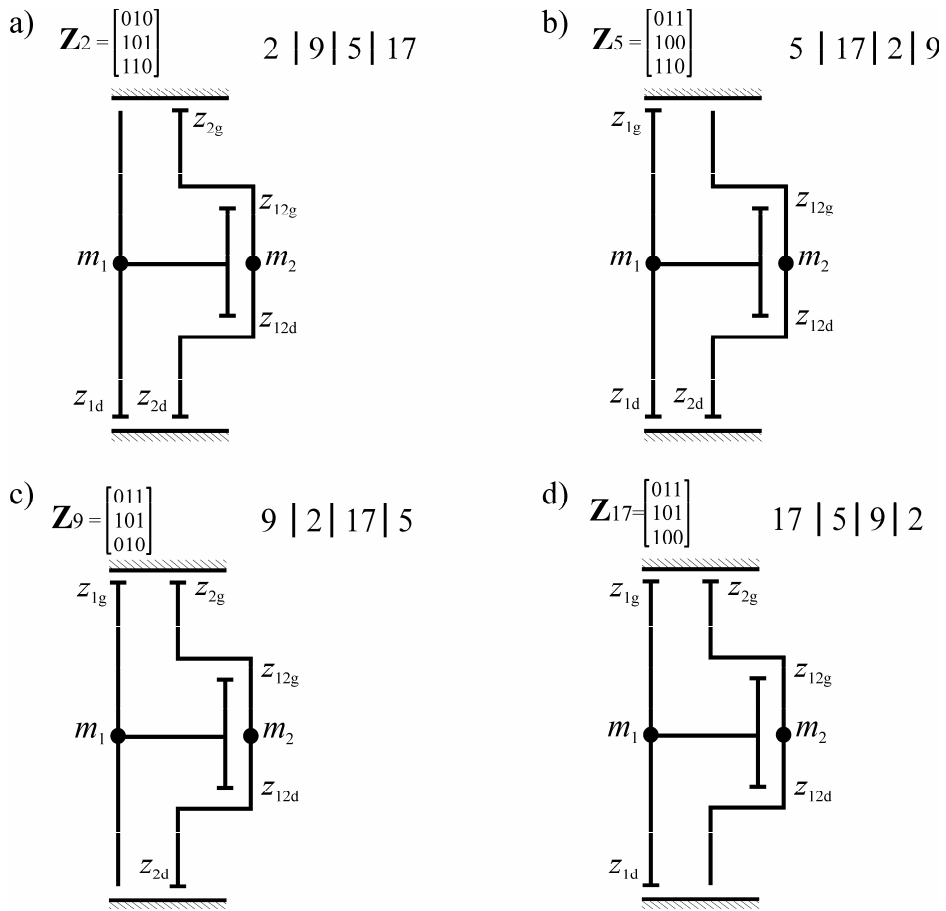

Fig. 6. Example of equivalent impact systems.

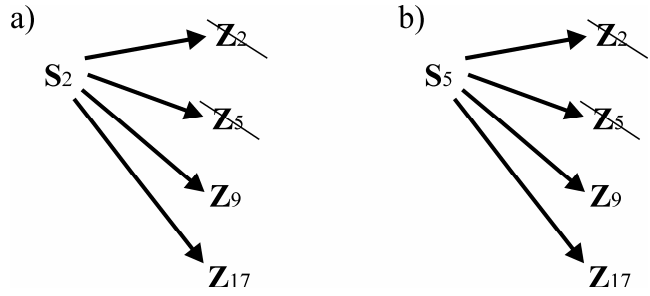

Fig. 7.   Scheme for matching the systems from Fig. 5 with the systems from Fig. 6.

*Description of the procedure of identification of equivalent systems*:  We assume that for the given degree of freedom *n*, we have all adjacency matrices of spring and impact systems (generated according to the procedure described in Section 4.1.1) at our disposal.

In the first phase, we will deal with impact systems. For each system, we find systems equivalent to it. The information on the kind of equivalency is recorded in the respective *impact information fields*: $\mathbf{Z}|\mathbf{Z}^T|\mathbf{Z}^P|\mathbf{Z}^{PT}$, where $\mathbf{Z}$ denotes the number of the matrix $\mathbf{Z}$ of the given system, and $\mathbf{Z}^T$, $\mathbf{Z}^P$, $\mathbf{Z}^{PT}$ - numbers of the matrices $\mathbf{Z}^T$, $\mathbf{Z}^P$, $\mathbf{Z}^{PT}$ of the systems equivalent to the given system via transposition, inversion and translocation, respectively. Thus, *a table of impact relations*, which includes full information on equivalence between impact systems, will be obtained.

In the second phase, we deal with spring systems. As each spring adjacency matrices $\mathbf{S}$ is symmetric, thus $\mathbf{S}^T=\mathbf{S}$ and $\mathbf{S}^{PT}=\mathbf{S}^P$. Hence, for each spring system, it is enough to find a system equivalent to it via inversion. The information on equivalency is written in two *fields of spring information*: $\mathbf{S}|\mathbf{S}^P$, where $\mathbf{S}$ denotes the number of matrix $\mathbf{S}$ of the given system, and $\mathbf{S}^P$ - the number of matrix $\mathbf{S}^P$ of the system equivalent to the given system via inversion. As a result, we will obtain *a table of spring relations* including information on equivalencies between spring systems.

In the last stage, we deal with spring-impact systems. We generate such systems by matching spring and impact systems with each other. Next, using the table of spring relations and the table of impact relations, we identify equivalent systems, according to principles (2), (3) and (4). Thus obtained sets of equivalent systems will be referred to as *classes of relations*.  Let us notice that undistinguishable systems as they correspond to one physical system, which has specified spring and impact connections of subsystems, belong to the same class of relations. Hence, it is necessary not only to identify all classes of relations, but to select representatives of classes and to eliminate the systems that are not representatives as well.

Below, *the principles for selection* of representatives of classes of relations are given, thus the criteria for elimination of equivalent spring-impact systems are specified. Let us remind that $\mathbf{S}$ and $\mathbf{S}^P$, and $\mathbf{Z}$, $\mathbf{Z}^T$, $\mathbf{Z}^P$ and $\mathbf{Z}^{PT}$ denote the fields of spring information and the fields of impact information, correspondingly.

**A selection of representatives of classes of relations of all spring-impact systems is conducted according to the following rules:**

1. The representatives of classes of relations will become the systems **SZ** obtained by matching spring systems **S** equivalent to themselves ($S=S^P$) with impact systems **Z** fulfilling the condition:

$$Z \geq Z^T \ \wedge \ Z \geq Z^P \ \wedge \ Z \geq Z^{PT} \tag{6}$$

    Hence, we eliminate the systems for which the equality $S=S^P$ holds and (6) does not hold.

2. The representatives of classes of relations are also the systems **SZ** obtained as a result of matching spring systems **S** that satisfy the condition $S>S^P$ with:

    **a)** impact systems **Z** that fulfill relation (6)

    and

    **b)** impact systems **Z** from the classes of relations of systems that satisfy the condition:

$$Z \geq Z^T \ \wedge \ Z > Z^P \ \wedge \ Z > Z^{PT} \tag{7}$$

    of the numbers equal to a higher number out of two numbers: $Z^P$ and $Z^{PT}$.

    We eliminate thus the systems that do not satisfy principle 2 and the systems that are generated by matching all spring systems fulfilling the condition $S<S^P$ with all impact systems.

The above-mentioned principles for selection of representatives of classes of spring-impact representatives need to be commented on.

In the first principle, spring systems that are equivalent to themselves via inversion are meant. Having matched such systems with impact systems that fulfill condition (6), we choose as the representatives of classes of relations these systems which have *the highest numeration*, i.e., among systems with the highest value of the number field of the spring matrix, we choose that one which has the highest value of the number field of the impact matrix.

The possible cases are as follows:

1) Matching the spring system equivalent to itself via inversion with four impact systems from one class of impact relations (a class of impact relations can include one, two, three or four elements) yields four equivalent spring-impact elements. We select one representative (with the highest numeration) among the four-element class, and we eliminate the remaining systems.

2) If we match the spring system equivalent to itself via inversion with impact systems from one class of impact relations, in which there are the following systems: symmetric (equivalent to itself via transposition) and equivalent:

a) to itself via inversion and via translocation, then there will be one spring-impact system in the class of relations, which will be the representative of the class of relation at the same time;

b) to another system via inversion and via translocation, then in the class of relations there will be two spring-impact systems, and the system of the highest numeration will be the representative of this class of relations.

3) If we match a spring system equivalent to itself via inversion with impact systems from one class of impact relations, in which there are the following systems: unsymmetrical (non-equivalent to itself via transposition) and equivalent to itself via inversion or via translocation, then there will be two spring-impact systems in the class of relations, and the system with the highest numeration will be the representative of this class of relation.

The second principle concerns spring systems whose numbers are higher than numbers in their fields of inversion ($S>S^P$). Having matched such systems with impact systems that satisfy conditions (6) and (7), we choose systems with the highest numeration as the representatives of classes of relations. Matches of the spring systems fulfilling the condition $S<S^P$ with the impact systems satisfying conditions (6) and (7) will never be the representatives of classes of relations. Let us consider the two spring systems shown in Fig. 5.a and Fig. 5.b. The spring adjacency matrices of these systems will be denoted as $S_2$ and $S_5$. Let us see that $S_2^P=S_5$, and thus the systems $S_2$ and $S_5$ are equivalent via inversion. As $2=S_2<S_2^P=5$, thus the system $S_5$ has a higher numeration and it is the representative of the two-element class of spring relations $\{S_2, S_5\}$ (classes of spring relations can have one or two elements).

Now, let us consider the four impact systems shown in Fig. 6.a, Fig. 6.b, Fig. 6.c and Fig. 6.d. The adjacency matrices of these systems will be denoted as $Z_2$, $Z_5$, $Z_9$ and $Z_{17}$, correspondingly. The systems are equivalent and they form a four-element class of impact relations $\{Z_2, Z_5, Z_9, Z_{17}\}$.

In Fig. 7.a a scheme of matching the system $S_2$ with all impact systems from Fig. 6 is shown (we will obtain four cases). Figure 7.b presents a scheme of matching the system $S_5$ with all impact systems from Fig. 6 (here we will receive four cases as well). Having in mind the fact that each spring adjacency matrix is symmetric, we should eliminate, via transposed equivalency, the following spring-impact systems (leaving the systems of a higher numeration): in Fig. 7.a - $S_2Z_2$ ($S_2Z_2^T \sim S_2Z_9$) and $S_2Z_5$ ($S_2Z_5^T \sim S_2Z_{17}$), and in Fig. 7.b - $S_5Z_2$ ($S_5Z_2^T \sim S_5Z_9$) and $S_5Z_5$ ($S_5Z_5^T \sim S_5Z_{17}$).

It can be stated that having applied the transposed equivalency, eight equivalent spring-impact systems are reduced to four systems, which are equivalent to the systems eliminated (matches crossed off in Fig. 7). By using the inversed and translocated equivalency, four nonelimianted systems can be reduced to two. In such a situation, we leave the spring-impact systems with the highest numeration: $S_5Z_{17}$ ($S_2Z_9^P \sim S_5Z_{17}$) and $S_5Z_9$ ($S_2Z_{17}^P \sim S_5Z_9$). Let us notice that the first system $S_5Z_{17}$ is the case described by principle 2a, i.e., a matching of the spring system that fulfills the condition $S>S^P$ with the impact system from the table of impact relations that fulfills condition (6). The second system $S_5Z_9$ is the case described with principle 2b, i.e., a matching of the spring system fulfilling the condition $S>S^P$ with the impact system from the class of relations of the system fulfilling condition (7) of the number higher out of two numbers $Z^P$ and $Z^{PT}$.

The following conclusions can be drawn from the above analysis, namely:

1) As a result of matching two spring systems equivalent via inversion with four impact systems from one class of impact relations, two classes of spring-impact relations arise – they both have four elements. In the above-described case, which has been selected among numerous cases of matches for $n=2$, there are the following systems in one class of spring-impact relations: $S_5Z_{17}$ (the representative of the class of relations), $S_2Z_2$, $S_2Z_9$, $S_5Z_5$. The second class of spring-impact relations comprises the following systems: $S_5Z_9$ (the representative of the class of relations), $S_5Z_2$, $S_2Z_5$, $S_2Z_{17}$. Let us notice that the representative $S_5Z_{17}$ has originated as a result of matching the spring system fulfilling the condition $S>S^P$ with the impact system satisfying condition (6), whereas the representative $S_5Z_9$ has come as a result of matching the spring system fulfilling the condition $S>S^P$ with the impact system satisfying condition (7).

2) If we match two spring systems equivalent to each other via inversion with two impact systems from one class of impact relations, in which there are the following systems: symmetrical (equivalent to itself via transposition) and equivalent

       a) to itself via inversion and translocation, then one class of spring-impact relations will arise and it will have two elements,

       b) to another system via inversion and translocation, then two classes of spring-impact relations will arise and each will have two elements.

3) If we match two spring systems equivalent via inversion with impact systems from one class of impact relations in which there are the systems: unsymmetrical (non-equivalent to itself via transposition) and equivalent to itself via inversion and translocation, then only one class of spring-impact relations will arise and it will have four elements.

Further on, we will consider representatives of spring-impact classes of relations only. Connected and disconnected systems are among them.

## 4. 2. Identification of connected systems

Subphase II of *Phase II* (Section 2) comprises elimination of all systems in which a subdivision of the spring-impact system into two or more independent systems, which are not connected either by a spring or a fender, has occurred. To identify these systems, we will use the notion of connectedness of the graph.

Let us notice that in the spring-impact system that represents a mechanical system with impacts, a subdivision into at least two independent systems will occur when the graph formed from the spring-impact graph as a result of neglecting the vertex corresponding to the frame and all the edges incidental to it, will not be connected. The analysis of the graph connectedness is conducted with the algorithm for integration of vertices (see Deo [20]). The basic step in this algorithm is an integration of the adjacent vertices. We start with a certain vertex in the graph and we integrate all the vertices adjacent to it. Then, we take the integrated vertex and again integrate it with all vertices that are now adjacent to it. The integration procedure continues until it is not possible to integrate any more vertices. It indicates that a certain connected component has been "integrated" to a single vertex. If it refers to all vertices in the graph, the graph is connected. Otherwise, we start from a certain new vertex (in another component) and we follow the integration procedure. In the adjacency matrix, the integration of the $j$th vertex with the $i$th vertex is done with the procedure OR, i.e., the logical addition of the $j$th row to the $i$th row and the $j$th column to the $i$th column. Let us keep in mind that in the logical addition $1+0=0+1=1+1=1$ and $0+0=0$.

The algorithm for integration of vertices can be used in investigations of connectedness of spring and impact systems. In the case of impact systems, one should remember that according to the connectedness definition of the directed graph, the adjacency matrix $\mathbf{Z}=[z_{ij}]$ of the system should be replaced by the matrix $\mathbf{Z'}=[z'_{ij}]$, where $z'_{ij}=z_{ij}$ OR $z_{ji}$ for all $i, j$. Such an operation will be referred to as *the matrix symmetrization*.

The investigations of the connectedness of the spring-impact system consists in the application of the above-mentioned algorithm to the matrix $\mathbf{M}=[m_{ij}]$, obtained as a result of the logical sum of the spring adjacency matrix $\mathbf{S}$ and the symmetrized impact adjacency matrix $\mathbf{Z}$ ($m_{ij}=s_{ij}$ OR $z_{ij}$ OR $z_{ji}$ for all $i, j$). In the considerations devoted to the connectedness of spring, impact or spring-impact systems, the last rows and the last columns of the adjacency matrix (connections of the subsystems with the frame and connections of the frame with the subsystems are of no significance here) can be neglected.

## 4. 3. Classification method

Employing all the above-described procedures, we identify equivalent combinations, select representatives of classes of relations and identify connected spring-impact systems. The so-obtained representatives of classes of spring-impact relations form a set of all structural patterns of vibro-impact systems with an arbitrary number of degrees of freedom. This is a consequence of the fact that instead of a spring connection, we can introduce any other connection that describes the action of at least one force (linear or nonlinear) that depends on displacement or velocity in the system. It can be an elasticity force, but also a viscous damping force, a friction force or an elastic-damping force, or even a triple combination of these forces. By a structural pattern of the technical system with impacts is understood a certain series of systems characterized by a specified structure of component elements (a definite configuration of fenders and connections). All structural patterns of mechanical systems constitute a set in which a kind of the connection (a spring or a damper) and its character (linearity or nonlinearity) and a way the impact phenomenon is modeled are not differentiating parameters.

The proposed classification of mechanical systems with impacts according to the characteristic properties of their structure allows us to rearrange the knowledge on systems with impacts and is the basis for understanding the sources of their diversity. Providing a full set of objects to be analyzed, it gives hints for new ideas and directions in designing mechanical devices. Obviously, it will not always satisfy fully designers for whom a functional classification allowing for a selection of the proper system functionally indispensable in the given device would be equally important. However, it has not been possible to combine the properties of structure and function in any existing classification yet. Besides, trials to develop a functional classification would be unsatisfactory due to two reasons. Firstly, one system can belong simultaneously to a few different classes considered in functional terms. Secondly, as the progress in technology goes further and further, new functions of systems with impacts can appear. Thus, in principle the functional classification would not fulfill the condition of exclusiveness and full completeness, which is satisfied by the structural classification presented herein.

The presented method of identification and description of structural patterns will be discussed in the author's next study on the example of systems with one and two degrees of freedom.

## 5. Conclusions

A remarkable increase in the interest in investigations of more and more complex mechanical systems with impacts, as well as a multitude and a diversity of such systems impose a need to classify them. Taking advantage of simplicity of the spring connection that commonly occurs in mechanical systems with impacts, a classification method for systems with an arbitrary number of degrees of freedom has been proposed.

The essence of the proposed method consists in a proper matching of spring and impact systems. Thus obtained systems (spring-impact systems) can be connected or disconnected. In the case of systems with two degrees of freedom, the matchings of disconnected spring and impact systems lead to disconnected spring-impact systems. However, for systems with three or more degrees of freedom, the situation does not to have to be the same. Therefore, while building more complex systems, disconnected spring and impact systems should be accounted for.

In the notation of relations occurring in vibro-impact systems, a certain matrix representation that allows for determination of all systems has been introduced. However, its disadvantage lies in the fact that it is possible to assign various adjacency matrices to the same physical system. To overcome this discrepancy, procedures for identification and elimination of unnecessary adjacency matrices have been developed. In the further considerations, only representatives of spring-impact class of relations are analyzed. The criteria for their selection are given in the selection principles developed. Next, a procedure for identification of connected and disconnected systems, i.e., systems in which a subdivision into independent systems does not occur or occurs, respectively, is introduced. The developed method enables structural classification of systems with impacts with an arbitrary number of degrees of freedom.

This study provides numerous data that extend the knowledge on mechanical systems with impacts. In future, this information can be used in designing such structures. The knowledge of properties of individual types of systems and principles of their formation can be helpful in solving various technical tasks that fall beyond the scope of traditional applications.

**References:**
[1] Awrejcewicz J., Lamarque C-H., Bifurcation and chaos in nonsmooth dynamical systems. World Scientific Series in Nonlinear Science, Series A. World Scientific Publishing Co. Singapore, 2003.
[2] Babitskii V.I., Theory of Vibroimpact Systems. Moscow: Nauka (in Russian), 1978.
[3] Babitsky V.I, Krupenin V.L., Vibrations in strongly nonlinear systems. Moscow: Nauka (in Russian), 1985.
[4] Bajkowski, J., Modeling of shock absorbers with dry friction subjected in impact loads. *Machine Dynamics Problems*, 1996, 16, pp. 7-19.
[5] Bapat C.N., Periodic motions of an impact oscillator. *Journal of Sound and Vibration*, 1998, 209(1), pp. 43-60.
[6] Bazhenov V.A., Pogorelova O.S., Postnikova T.G., Lukyanchenko O.A., Numerical investigations of the dynamic processes in vibroimpact systems in modeling impacts by a force of contact interaction. *Strength of Materials*, 2008, 40, pp. 656-662.
[7] Blazejczyk-Okolewska B., Kapitaniak T., Dynamics of Impact Oscillator with Dry Friction. *Chaos, Solitons & Fractals*, 1996, 7(1), pp. 1-5.
[8] Blazejczyk-Okolewska B., Kapitaniak T., Co-existing attractors of impact oscillator. *Chaos, Solitons & Fractals*, 1998, 9(8), pp. 1439-1443.
[9] Blazejczyk-Okolewska B., Czolczynski K., Kapitaniak K., Classification principles of types of mechanical systems with impacts - fundamental assumptions and rules. *European Journal of Mechanics A/Solids*, 2004, 23, pp. 517-537.
[10] Blazejczyk-Okolewska B., Czolczynski K., Kapitaniak T., Wojewoda J., Chaotic Mechanics in

Systems with Impacts and Friction. World Scientific Series in Nonlinear Science, Series A. World Scientific Publishing Co. Singapore, 1999.
[11] Brach R.M., Mechanical Impact Dynamics. Rigid Body Collisions. John Wiley and Sons, Inc., 1991.
[12] Brogliato B., Nonsmooth Mechanics. Springer, 1999.
[13] Cempel C., The periodical vibration with impacts in mechanical discrete systems. Poznan Technical University, Poznan, Scientific Dissertations No. 44 (in Polish), 1970.
[14] Cempel C., The multi-unit impact damper: equivalent continuous force approach. *Journal of Sound and Vibration*, 1974, 34(2), pp. 199-209.
[15] Chin W., Ott E., Nusse H.E., Grebogi C., Grazing bifurcation in impact oscillators. *Physical Review E*, 1994, 50(6), pp. 4427-4444..
[16] Czolczynski K., Kapitaniak T., Influence of the mass and stiffness ratio on a periodic motion of two impacting oscillators. *Chaos, Solitons & Fractals*, 2003, 17, pp. 1-10.
[17] de Souza S.L.T., Caldas I.L., Calculation of Lyapunov exponents in systems with impacts. *Chaos, Solitons & Fractals*, 2004, 19, pp. 569-579.
[18] de Souza S.L.T., Batista A.M., Caldas I.L., Viana R.L., Kapitaniak T., Noise-induced basin hopping in a vibro-impact system. *Chaos, Solitons & Fractals*, 2007, 32, pp. 758-767.
[19] Dabrowski A., Kapitaniak T., Using chaos to reduce oscillations: Experimental results. *Chaos, Solitons & Fractals*, 2009, 39(4), pp. 1677-1683.
[20] Deo N., Graph Theory with Applications to Engineering and Computer Science, Prentice-Hall, Inc., Englewood Cliffs, N. J., 1974.
[21] Du Y., Wang S., Modeling the fine particle impact damper. *International Journal of Mechanical Sciences*, 2010, 52, pp. 1015-1022.
[22] Goldsmith W., Impact. Arnold LTD, London, 1960.
[23] Hinrichs N., Oestreich M., Popp K., Dynamics of oscillators with impact and friction. *Chaos, Solitons & Fractals*, 1997, 8(4), pp. 535-558.
[24] Ho J.-H., Nguyen V.-D., Woo K.-C., Nonlinear dynamics of a new electro-vibro-impact system. *Nonlinear Dynamic*, 2011, 63, pp. 35-49.
[25] Ibrahim R.A., Vibro-impact Dynamics: Modeling, Mapping and Applications. Lecture Notes in Applied and Computational Mechanics. Springer-Verlag Berlin Heidelberg, 2009.
[26] Kaharaman A., Singh R., Nonlinear Dynamics of a Spur Gear Pair. *Journal of Sound and Vibration*, 1990, 142, pp. 49-75.
[27] Kobrinskii A.E., Kobrinskii A., Vibroimpact System. (In Russian), Moscow: Nauka Press, 1973.
[28] Lee Y.S., Nucera F., Vakakis A., McFarland M., Periodic orbits, damped transitions and targeted energy transfers in oscillators with vibro-impact attachments. *Physica D*, 2009, 238, pp. 1868-1896.
[29] Lin S.Q., Bapat C.N., Estimation of clearances and impact forces using vibroimpact response: random excitation. *Journal of Sound and Vibration*, 1993, 163, pp. 407-421.
[30] Luo A.C.J., Aperiodically forced, piecewise linear system. Part I: Local singularity and grazing bifurcation. *Communication in Nonlinear Science and Numerical Simulation*, 2007, 12, pp. 379-396.
[31] Luo G.W., Lv X.H., Controlling bifurcation and chaos of a plastic impact oscillator. *Nonlinear Analysis: Real World Applications*, 2009, 10, pp. 2047-2061.
[32] Maistrenko Y., Kapitaniak T., Szuminski P., Locally and globally riddled basins in two coupled piecewise-linear maps. *Physical Review E*, 1997, 56 (6), pp. 6393-6399.
[33] Masri S.F., Ibrahim A.M., Stochastic excitation of a simple system with impact damper. *Earthquake Engineering and Structural Dynamics*, 1973, 1, pp. 337-346.
[34] Miandoab E.E.M., Yousefi-Koma A., Ehyaei D., Optimal Design of an impact damper for a nonlinear friction-driven oscillator. *International Journal of Mathematical Models and Methods in Applied Sciences*, 2008, 2(2), 236-243.

[35] Natsiavas S., Dynamics of multiple-degree-of-freedom oscillators with colliding components. *Journal of Sound and Vibration*, 1993, 165, pp. 439-453.

[36] Nguyen D.T., Noah S.T., Kettleborough C.F., Impact behavior of an oscillator with limiting stops (part I and II). *Journal of Sound and Vibration*, 1987, 109, pp. 293-325.

[37] Nigm M.M, Shabana A.A, Effects of an impact damper on a multidegree of freedom system. *Journal of Sound and Vibration*, 1983, 89, pp. 541-557.

[38] Nordmark A.B., Non-Periodic Motion Caused by Grazing Incidence in an Impact Oscillator. *Journal of Sound and Vibration*, 1991, 145(2), pp. 279-297.

[39] Peterka F., An investigation of the motion of impact dampers, paper I, II, III. Strojnicku Casopis XXI, c.5 (in Czech Republic), 1971.

[40] Peterka F., Introduction to vibration of mechanical systems with internal impacts. Academia. Praha, 1981.

[41] Peterka F., Blazejczyk-Okolewska B., Some Aspects of the Dynamical Behavior of the Impact Damper. *Journal of Vibration and Control*, 2005, 11(4), pp. 459-479.

[42] Shaw S.W., Holmes P.J., A Periodically Forced Piecewise Linear Oscillator. *Journal of Sound and Vibration*, 1983, 90(1), pp. 129-155.

[43] Stefanski A., Kapitaniak T., Using chaos synchronization to estimate the largest Lyapunov exponent of nonsmooth systems. *Discrete dynamics in Nature and Society*, 2000, 4(3), pp. 207-215.

[44] Stefanski A., Kapitaniak T., Estimation of the dominant Lyapunov exponent of non-smooth systems on the basis of maps synchronization. *Chaos, Solitons & Fractals*, 2003, 15(2), pp. 233-244.

[45] Tung P.C., Shaw S.W., The Dynamics of an Impact Print Hammer. *ASME J. Vibration Stress Reliability in Design*, 1988, 110, pp. 193-199.

[46] Wiercigroch M., Modelling of dynamical systems with motion dependent discontinuities. *Chaos, Solitons & Fractals*, 2000, 11, pp. 2429-2442.